\documentclass[fleqn,usenatbib]{mnras}

\usepackage[T1]{fontenc}
\usepackage{ae,aecompl}

\usepackage{graphicx}	
\usepackage{amsmath}	
\usepackage{threeparttable} 
\usepackage{booktabs,siunitx} 
\usepackage{array,tabularx,ragged2e}
\usepackage{pdflscape}
\usepackage{xcolor, soul}
\usepackage{adjustbox}
\usepackage{fnpct} 

\newcommand{\teff}{$T_{\mathrm{eff}}$}
\newcommand{\muhz}{$\mathrm{\mu}$Hz}
\newcommand{\numax}{$\mathrm{\nu}_{\mathrm{max}}$}
\newcommand{\dnu}{$\mathrm{\Delta\nu}$}
\newcommand{\logg}{$\log g$}

\newcommand{\msun}{M$_\mathrm{\odot}$}
\newcommand{\rsun}{R$_\mathrm{\odot}$}

\newcommand{\kepler}{\textit{Kepler}}

\newcommand{\echelle}{\'echelle}

\newcommand{\tess}{TESS}
\newcommand{\lsun}{$\mathrm{L_{\sun}}$}

\newcommand{\mnumax}{M$_\mathrm{1}$}

\newcolumntype{P}[1]{>{\RaggedRight\hspace{0pt}}p{#1}} 

\title[Benchmarking the spectroscopic masses of 249 evolved stars]{Benchmarking the spectroscopic masses of 249 evolved stars using asteroseismology with TESS}

\author[Malla et al.]{
Sai Prathyusha Malla$^{1}$\thanks{E-mail: s.malla@student.unsw.edu.au (UNSW)},
Dennis Stello$^{1, 2}$,
Benjamin T. Montet$^{1}$,
Daniel Huber$^{3}$,
\newauthor
Marc Hon$^{1, 3}$,
Timothy R. Bedding$^{2}$,
Claudia Reyes$^{1}$, and 
Daniel R. Hey$^{2,3}$
\\
$^{1}$School of Physics, The University of New South Wales, Sydney NSW 2052, Australia\\
$^{2}$Sydney Institute of Astronomy(SIfA), School of Physics, University of Sydney, NSW 2006, Australia\\
$^{3}$Institute for Astronomy, University of Hawai`i, 2680 Woodlawn Drive, Honolulu, HI 96822, USA\\
}

\date{Accepted XXX. Received YYY; in original form ZZZ}

\pubyear{2024}

\usepackage{newtxtext,newtxmath}
\begin{document}
\label{firstpage}
\pagerange{\pageref{firstpage}--\pageref{lastpage}}
\maketitle

\begin{abstract}

One way to understand planet formation is through studying the correlations between planet occurrence rates and stellar mass. However, measuring stellar mass in the red giant regime is very difficult. In particular, the spectroscopic masses of certain evolved stars, often referred to as "retired A-stars", have been questioned in the literature. Efforts to resolve this mass controversy using spectroscopy, interferometry and asteroseismology have so far been inconclusive. A recent ensemble study found a mass-dependent mass offset, but the result was based on only 16 stars. With NASA's Transiting Exoplanet Survey Satellite (TESS), we expand the investigation of the mass discrepancy to a total of 92 low-luminosity stars, synonymous with the retired A-stars. We measure their characteristic oscillation frequency, \numax, and the large frequency separation, \dnu, from their TESS photometric time series. Using these measurements and asteroseismic scaling relations, we derive asteroseismic masses and compare them with spectroscopic masses from five surveys, to comprehensively study the alleged mass-dependent mass offset. We find a mass offset between spectroscopy and seismology that increases with stellar mass. However, we note that adopting the seismic mass scale does not have a significant effect on the planet occurrence-mass-metallicity correlation for the so-called retired A-stars. We also report seismic measurements and masses for 157 higher luminosity giants (mostly helium-core-burning) from the spectroscopic surveys.
\end{abstract}




\section{Introduction}
Correlations between the planet occurrence rate and host star properties, in particular stellar mass, can inform us about planet formation (e.g., see \citealt{fischer_valenti_2005}, \citealt{johnson_giant_2010}, \citealt{dawson_hotjupiters_2005}, \citealt{ghezzi_retired_2018}, \citealt{yang_omm_2020} and \citealt{wolthoff_omm_2022}). For this, it is important to sample stars across a large range of masses, including those significantly more massive than the Sun. However, the main sequence A- and hot F-type stars have broad spectral lines making radial velocity measurements less precise. Hence, radial velocity planet searches have been biased against intermediate-mass A- and hot F- type stars. To circumvent this difficulty in finding planets around the intermediate-mass A- and hot F-type stars, \citet{johnson_eccentric_2006} set out to find planets around their cooler, more evolved counterparts in the subgiant and red giant regimes. They referred to these stars as "retired A"-stars. However, A- and F-progenitors that ascend the red giant branch become difficult to distinguish from their G- and K- counterparts because of their similar temperatures (\teff) and luminosities (L), leaving stellar mass as the only way to differentiate between these stars. Thus, accurately determining the stellar mass of these evolved stars is important. 

Traditionally, stellar masses are determined using spectroscopy and grid-based modelling via isochrone fitting. \citet{lloyd_retired_2011} compared the masses of a selection of evolved planet-hosting stars from the Exoplanet Orbit Database\footnote{\url{www.exoplanets.org}} \citep{wright_exoplanet_2011} with a sample of field stars with similar mass ($M$) and surface gravity (\logg) from the catalogs by \citet{prieto_fundamental_1999} and \citet{glebocki_rv_2005}. \citet{lloyd_retired_2011} discovered a discrepancy between the mass distributions for the samples of evolved planet-hosting stars from the Exoplanet Orbit Database and the field stars and suggested the masses of these evolved planet-hosting stars were overestimated by up to 50\%. They attributed the deficit of massive evolved planet-hosting stars in their sample to the differences in evolution speeds of stars with different stellar masses. Massive stars evolve faster than less massive stars and are hence less common, and the failure to account for this difference in evolution speeds of stars in grid-based modelling may lead to erroneous determination of the stellar mass. 

Later, \citet{johnson_retired_2013} benchmarked their apparent-magnitude-limited subgiant sample against Galactic stellar population models by \citet{girardi_trilegal_2005} and found no such overestimation in spectroscopic masses. \citet{johnson_retired_2013} argued that their imposed apparent magnitude limit partially counteracted the lower number of massive stars expected from their differential evolution by increasing the relative number of massive stars. However, \citet{lloyd_mass_2013} used apparent magnitude-limited weights for the isochrone integration in his 2011 calculations and argued that irrespective of such limit used in target selection, there was a deficit of massive stars in the subgiant and red giant regime compared to what was reported in the literature. Supporting this argument by \citet{lloyd_mass_2013}, \citet{schlaufman_evidence_2013} found consistency between the velocity dispersions for their subgiant sample and their main-sequence F5-G5 sample, but not with their main-sequence A0-F5 sample. This consistency led them to conclude that their subgiants are less massive than main sequence A0-F5 stars. 

In addition to the spectroscopy-based mass determinations, the retired A-star mass controversy has been explored using other methods like asteroseismology and interferometry. Asteroseismology, in particular, can provide us with precise model-independent stellar masses \citep{stello_oscillating_2008, kallinger_oscillating_2010, chaplin_asteroseismology_2013, gaulme_testing_2016, huber_asteroseismology_2017, yu2018asteroseismology} while long-baseline interferometry provides better constraints on effective temperatures and radii, thus reducing systematic errors in model-dependent stellar mass determinations \citep{timwhite_interferometry_2018}. \citet{johnson_physical_2014} made the first attempt to use asteroseismology to test the spectroscopic mass of HD 185351--the only known intermediate-mass evolved planet-hosting star observed by \kepler\ \citep{borucki_kepler_2010}. However, they could not reconcile their stellar mass measurements from spectroscopy, asteroseismology and interferometry. The reconciliation was later achieved by \citet{hjorringgaard_testing_2017} with a more comprehensive asteroseismic modelling. They found an overestimation of about $\sim$ 15\% in the spectroscopic mass of HD 185351. Similarly, for seven of the eight evolved planet-hosting stars observed using the ground-based Stellar Observations Network Group (SONG, \citealt{mads_song_2016}) telescope, \citet{stello_asteroseismic_2017} found an overestimation of 15--20\% in spectroscopic masses compared to the corresponding seismic masses.
Additionally, using interferometry for five evolved planet-hosting stars, \citet{timwhite_interferometry_2018} found the spectroscopic masses from the literature to be 15\% larger than their interferometry-based values. On the other hand, contrary to the above evidence suggesting the presence of a mass offset, \citet[1 star]{campante_weighing_2017} and \citet[7 stars]{north_masses_2017} found no apparent difference between the spectroscopic and seismic masses in their samples of stars, of which not all were hosting planets. \citet{ghezzi_parsec_2015} investigated the presence of a mass offset with 59 evolved stars using model-independent masses from binary systems (26 stars) and asteroseismology (33 stars) and found no significant evidence for it. Later, \citet[245 stars]{ghezzi_retired_2018} arrived at the same conclusion from their spectroscopic and kinematic analyses.

\citet{malla_retdA_2020} performed an ensemble study to understand why some seismic studies found a discrepancy between the spectroscopic and seismic mass scales while others did not. They combined their seismic results for four evolved planet-hosting stars with 12 other stars studied by \citet{stello_asteroseismic_2017} and \citet{north_masses_2017}. From three different spectroscopic sources \citep{mortier_new_2013,jofre_stellar_2015,stock_evoltime_2018}, they found that stars above a mass threshold of 1.6 \msun\ had a significant mass offset, while those below the threshold did not. This result is consistent with the findings by \citet{stello_asteroseismic_2017} and \citet{north_masses_2017}. The transition mass of $\sim$ 1.6 \msun\ is the same as the one that separates fast and slow-moving evolving stars in the subgiant regime where most of the retired A-stars lie. However, the \citet{malla_retdA_2020} ensemble study only had relatively few stars with masses below 1.0 \msun\ and above 1.6 \msun, leading to an inconclusive result. More stars across a wider mass range would be required to make firmer conclusions about such mass-dependent offset. 
Fortunately, with the launch of the Transiting Exoplanet Survey Satellite (TESS, \citealt{ricker_tess_2015}), we now have high-cadence time-resolved data for many more stars with spectroscopically-derived masses. Hence, we can use TESS to measure the asteroseismic-based masses of these stars. 

In this paper, we expand the sample of evolved stars for which we can test the previous spectroscopic mass determinations against seismically determined masses, and to provide more stars in the mass range poorly covered by \citet{malla_retdA_2020}. For this purpose, we use the photometric data from the first three years of the TESS mission and asteroseismic scaling relations to determine the seismic masses of these stars. With this large data set, we further expand the scope of the mass offset found by \citet{malla_retdA_2020} towards up to five different spectroscopic sources. In Section~\ref{secn:tarsel}, we discuss the selection criteria for our targets. We measure our global seismic parameters in Section~\ref{secn:seismo} and our seismic masses in Section~\ref{secn:seis_mass}. In Section~\ref{secn:retdA}, we discuss the mass discrepancy between spectroscopic and seismic mass scales guided by our results. In Section~\ref{secn:conclusions}, we draw conclusions based on our findings. 

\section{Target Selection and Observations} \label{secn:tarsel}

\begin{figure*}
	\includegraphics[width=\textwidth]{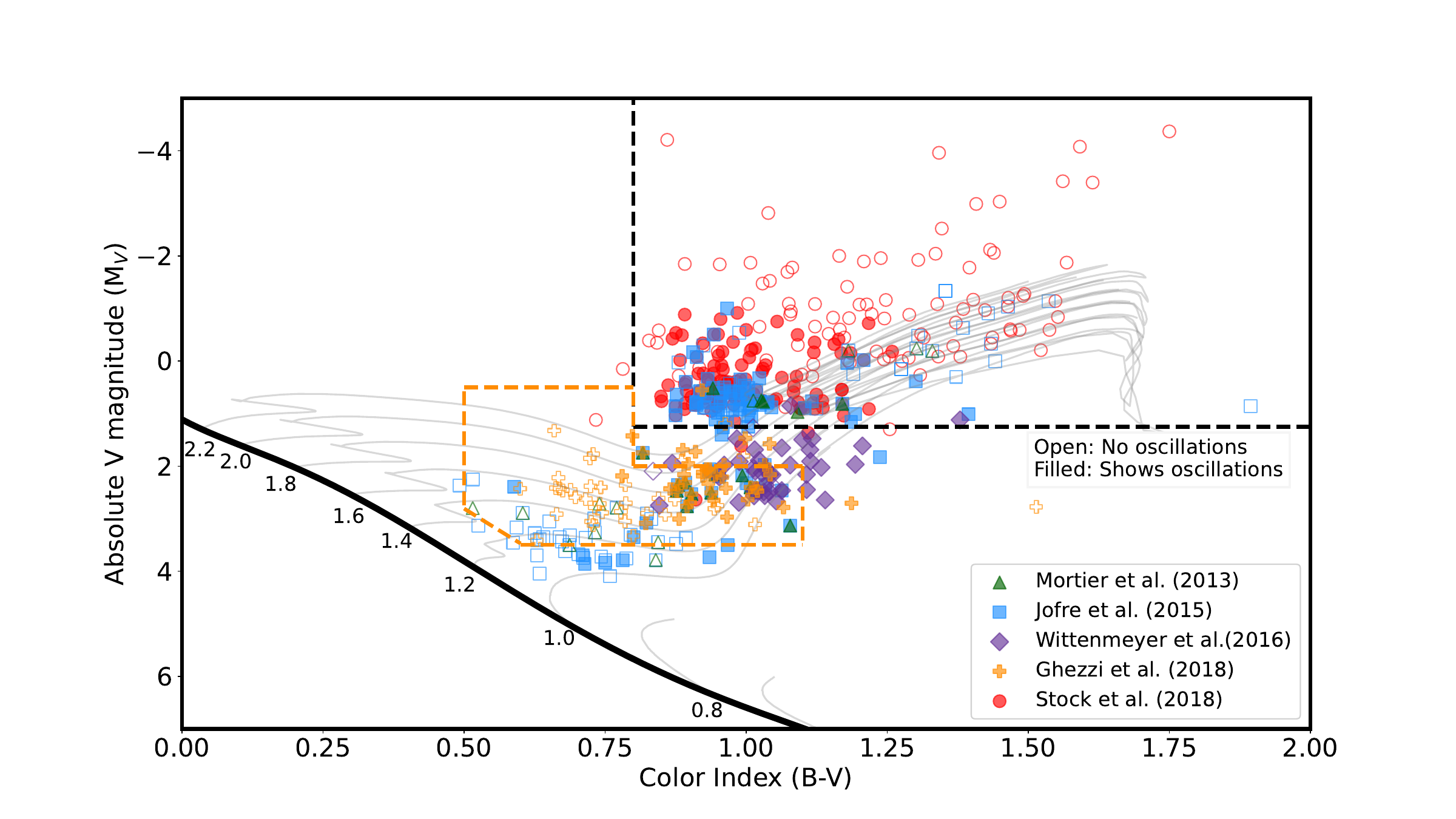}
    \caption{
     Color-magnitude diagram of the 451 stars in our sample plotted along the solar-metallicity BaSTI evolutionary tracks \citep{basti_pietrinferni_2004}. Each point represents a star and is colour-coded according to the spectroscopic source to which it belongs. Green triangles represent the stars previously studied by M13, blue squares by J15, purple diamonds by W16, orange pluses by G18 and red circles by S18. Stars that were studied by multiple spectroscopic sources are plotted multiple times. A list of the overlapping stars can be inferred from Table~\ref{table2:spec_masses} (Columns 6--10). The solid black line is the averaged Hipparcos main-sequence defined by \citet{wright_ccps_2005}. The dashed black line corresponds to the boundary between the low- and high- luminosity stars in the CMD according to the criteria listed in Section~\ref{secn:retdA}. The area surrounded by dashed orange lines corresponds to the initial selection criteria employed by G18. For stars TIC 356000102, TIC 258873063, TIC 471011913 and TIC 471012056, B magnitude values were not available from the Tess Input Catalog and hence, the B$_{T}$ and V$_{T}$ values from the Hipparcos catalog \citep{hog_tycho-2_nodate} were used. 
     }
    \label{fig:cmd}
\end{figure*}

\begin{figure}
	\includegraphics[width=\columnwidth, trim={0 6cm 0 0}, clip, keepaspectratio]{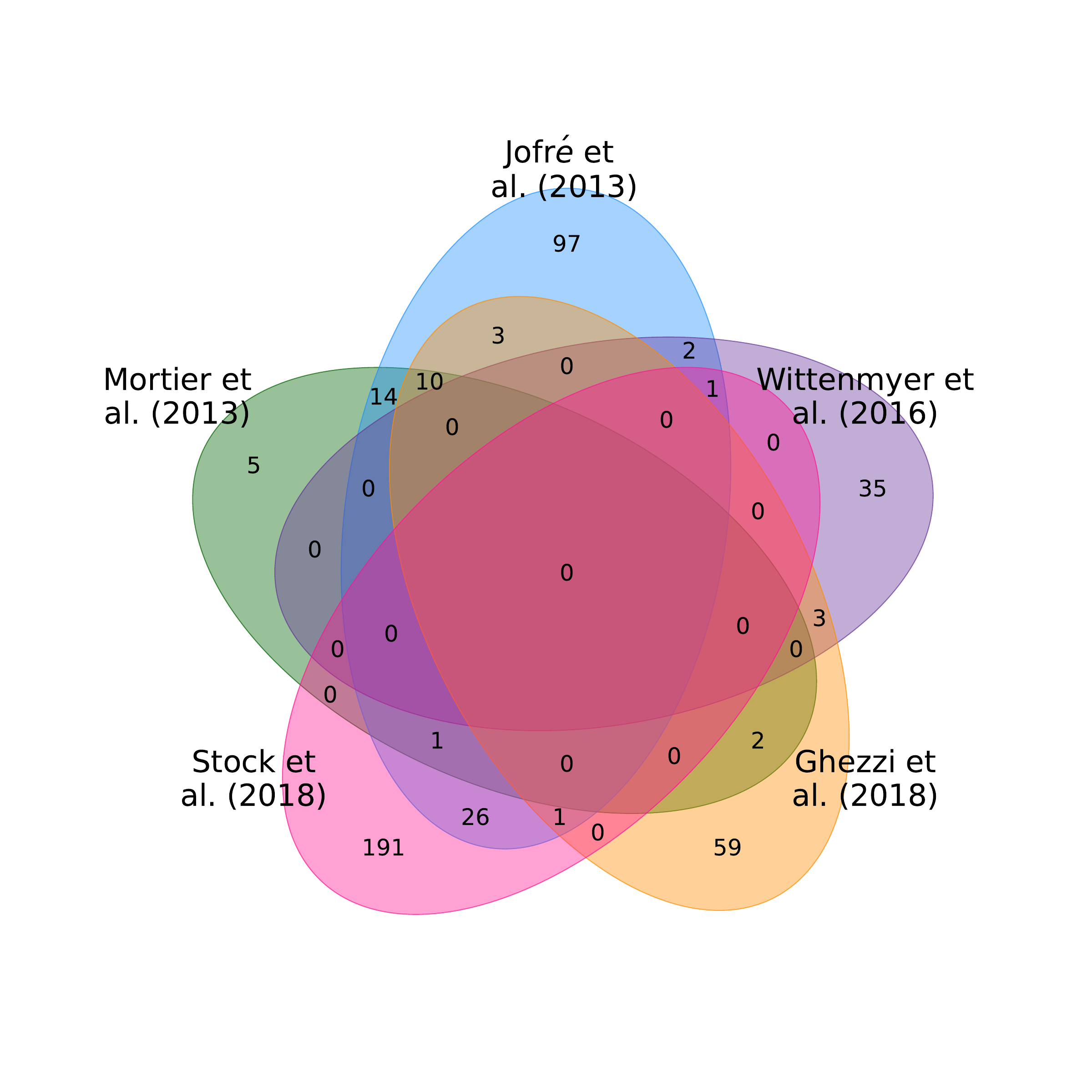}
    \caption{
     Venn diagram representing the 451 stars studied by the five spectroscopic sources: \citet[M13]{mortier_new_2013}, \citet[J15]{jofre_stellar_2015}, \citet[W16]{ppps}, \citet[G18]{ghezzi_retired_2018} and \citet[S18]{stock_evoltime_2018}. 
     }
    
    \label{fig:venn}
\end{figure}

We selected 451 evolved stars that had spectroscopic masses previously studied by either \citet[M13]{mortier_new_2013}, \citet[J15]{jofre_stellar_2015}, \citet[W16]{ppps}, \citet[G18]{ghezzi_retired_2018}, or by \citet[S18]{stock_evoltime_2018} and that were observed in 2-minute cadence during the first three years of the \tess\ mission. Table~\ref{table1:osci_stats} (Column 2) shows the total number of unique stars in our sample, along with the number of stars analyzed from each spectroscopic survey. All five spectroscopic sources derived their masses by fitting isochrones or evolutionary tracks, with J15 and S18\footnote{S18 provided spectroscopic masses based on the location of the star on the Hertzsprung-Russell diagram. They gave two masses depending on whether the star is on the red giant branch or the horizontal branch, along with an associated probability for each. For our purpose, similar to S18, we only chose the spectroscopic mass with the associated probability greater than 50\%.} utilizing Bayesian inference. We note here that S18 incorporated evolutionary speeds into their spectroscopic mass determinations. Fig.~\ref{fig:cmd} shows the colour-magnitude diagram for the 451 stars in our sample along with solar-metallicity BaSTI tracks \footnote{\url{http://basti.oa-teramo.inaf.it/}} \citep{basti_pietrinferni_2004} to guide the eye. Of the five chosen spectroscopic sources, M13, J15, and S18 studied evolved stars and were previously used to study the mass offset by \citet{malla_retdA_2020}. G18 studied evolved stars (both planet-hosting and non-planet-hosting) to derive a planet occurrence-mass-metallicity correlation, while W16 studied the retired A-stars as observed from the southern hemisphere. In Table~\ref{table1.5:spec_mass_range}, we list the spectroscopic mass range covered by our sample from each source and in Table~\ref{table2:spec_masses}, we list the spectroscopic masses of the stars for each spectroscopic source. From Fig.~\ref{fig:venn} and also Table~\ref{table2:spec_masses} (Columns 6--10), we can see that 66 stars are studied by multiple spectroscopic sources. Such stars are plotted multiple times in Fig.~\ref{fig:cmd}. 

\noindent\setlength\tabcolsep{2pt}%
\begin{table}
	\centering
	\caption{Star counts for TESS-spectroscopic overlap sample }
	\label{table1:osci_stats}
	\begin{threeparttable}
	
	\begin{tabularx}{\columnwidth}{P{1.4cm}X*{6}{>{\centering\arraybackslash}X}} 
		\toprule
		Spectroscopic source & $N_\mathrm{obs}$ & $N_\mathrm{osc}$ & $N_\mathrm{obs, low}$ & $N_\mathrm{osc, low}$ & $N_\mathrm{obs, high}$ & $N_\mathrm{osc, high}$ \\
         (1) & (2) & (3) & (4) & (5) & (6) & (7) \\
		\midrule
\textbf{Total unique stars} & 451 & 249 & 168 & 92 & 283 & 157 \\
\midrule
M13 & 32 & 19 & 22 & 14 & 10 & 5 \\
J15 & 162 & 92 & 66 & 31 & 96 & 61 \\
W16 & 41 & 38 & 39 & 36 & 2 & 2 \\
G18 & 78 & 41 & 76 & 40 & 2 & 1 \\
S18 & 220 & 116 & 8 & 5 & 212 & 111 \\
    	\bottomrule
    \end{tabularx}
      \begin{tablenotes}
        \item Notes: 
        \item $N_\mathrm{obs}$: Number of stars observed by TESS
        \item $N_\mathrm{osc}$: Number of oscillating stars
        \item $N_\mathrm{obs, low}$: Number of low-luminosity stars observed by TESS 
        \item $N_\mathrm{osc, low}$: Number of oscillating low-luminosity stars 
        \item $N_\mathrm{obs, high}$: Number of high-luminosity stars observed by TESS 
        \item $N_\mathrm{osc, high}$: Number of oscillating high-luminosity stars
        \item M13 : \citet{mortier_new_2013}
        \item J15 : \citet{jofre_stellar_2015}
        \item W16 : \citet{ppps}
        \item G18 : \citet{ghezzi_retired_2018}
        \item S18 : \citet{stock_evoltime_2018}
        \item A machine-readable version of the table is available online on CDS.
    \end{tablenotes}
	\end{threeparttable}
\end{table}

\noindent\setlength\tabcolsep{2pt}%
\begin{table*}
	\centering
	\caption{Spectroscopic mass range covered for TESS-spectroscopic overlap sample }
	\label{table1.5:spec_mass_range}
	\begin{threeparttable}
	
	\begin{tabularx}{\textwidth}{lc*{6}{>{\centering\arraybackslash}X}} 
		\toprule
		Spectroscopic source & M$_\mathrm{spec} $ & M$_\mathrm{spec, osc}$& M$_\mathrm{spec}^\mathrm{low}$  & M$_\mathrm{spec, osc}^\mathrm{low}$ & M$_\mathrm{spec}^\mathrm{high}$ & M$_\mathrm{spec, osc}^\mathrm{high}$ \\
         (1) & (2) & (3) & (4) & (5) & (6) & (7) \\
		\midrule
M13 & 0.98--2.31 & 1.04--2.31 & 1.08--2.03 & 1.24--2.03 & 0.98--2.31 & 1.04--2.31 \\
J15 & 0.74--3.33 & 0.91--3.23 & 0.91--1.99 & 0.91--1.99 & 0.74--3.33 & 1.01--3.23 \\
W16 & 1.03--2.19 & 1.03--2.19 & 1.03--2.19 & 1.03--2.19 & 1.62--1.66 & 1.62--1.66 \\
G18 & 0.92--2.08 & 0.92--2.08 & 0.92--2.08 & 0.92--2.08 & 1.33--1.86 & 1.33--1.33 \\
S18 & 0.84--6.65 & 0.84--3.23 & 0.96--2.29 & 0.96--1.69 & 0.84--6.65 & 0.84--3.23 \\

        
    	\bottomrule
    \end{tabularx}
      \begin{tablenotes}
        \item {Notes: \item A machine-readable version of the table is available online on CDS.}
        \item M13 : \citet{mortier_new_2013}
        \item J15 : \citet{jofre_stellar_2015}
        \item W16 : \citet{ppps}
        \item G18 : \citet{ghezzi_retired_2018}
        \item S18 : \citet{stock_evoltime_2018}
        \item M$_\mathrm{spec}$: Spectroscopic mass range for all 451 stars
        \item M$_\mathrm{spec, osc}$ : Spectroscopic mass for all 249 oscillating stars
        \item M$_\mathrm{spec}^\mathrm{low}$ : Spectroscopic mass range for all 168 low luminosity stars
        \item M$_\mathrm{spec, osc}^\mathrm{low}$ : Spectroscopic mass range for all 92 oscillating low luminosity stars
        \item M$_\mathrm{spec}^\mathrm{high}$ : Spectroscopic mass range for all 283 high luminosity stars
        \item M$_\mathrm{spec, osc}^\mathrm{high}$ : Spectroscopic mass range for all 157 oscillating high luminosity stars
    \end{tablenotes}
	\end{threeparttable}
\end{table*}

\noindent\setlength\tabcolsep{2pt}%
\begin{table*}
	\centering
	\caption{TESS coverage and spectroscopic masses of the targets}
	\label{table2:spec_masses}
	\begin{threeparttable}
	
	\begin{tabularx}{\textwidth}{P{1.4cm}P{1.4cm}*{9}{>{\centering\arraybackslash}X}} 
		\toprule
		TIC ID & HD & Name & TESS Magnitude & Number of sectors used &M$_\mathrm{M13}$ & M$_\mathrm{J15}$ & M$_\mathrm{W16}$& M$_\mathrm{G18}$ & M$_\mathrm{S18}$ & Category\\
		& & & & &\msun\ & \msun\ & \msun\ & \msun\ & \msun\ & \\
		(1) & (2) & (3) & (4) & (5) &  (6)\tnote{a} & (7)\tnote{b} &(8)\tnote{c} & (9)\tnote{d} & (10)\tnote{e} & (11)\tnote{f}\\
		\midrule
612908 & 30856 & - & 7.08 & 2 & 1.36 $\pm$ 0.07 & 1.31 $\pm$ 0.11 & - & 1.37 $\pm$ 0.07 & - & L \\
1713457 & 90043 &  24 Sex  & 5.64 & 2 & 1.81 $\pm$ 0.08 & 1.78 $\pm$ 0.08 & - & 1.71 $\pm$ 0.04 & - & L \\
5630694 & 103616 & - & 6.9 & 1 & - & - & - & 1.58 $\pm$ 0.07 & - & L \\
6013637 & 115659 &  46 Hya  & 2.17 & 2 & - & 3.09 $\pm$ 0.05 & - & - & 2.71 $\pm$ 0.19 & H \\
9710105 & 223807 & - & 4.76 & 2 & - & - & - & - & 2.52 $\pm$ 0.8 & H \\
11557059 & 7931 & - & 7.01 & 1 & - & - & 1.12 $\pm$ 0.25 & 1.3 $\pm$ 0.08 & - & L \\
29921672 & 10011 & - & 7.2 & 1 & - & - & - & 1.54 $\pm$ 0.09 & - & L \\
32550970 & 8407 & - & 6.92 & 1 & - & - & - & 1.13 $\pm$ 0.06 & - & L \\
47336943 & 215049 & - & 7.5 & 1 & - & - & - & 1.42 $\pm$ 0.09 & - & L \\
48401372 & 47562 & - & 7.45 & 2 & - & - & - & 1.63 $\pm$ 0.23 & - & L \\
48777312 & 190647 & - & 7.13 & 1 & - & 1.02 $\pm$ 0.02 & - & - & - & L \\
53873088 & 5608 & - & 5.14 & 1 & 1.66 $\pm$ 0.08 & 1.72 $\pm$ 0.07 & - & - & - & L \\
141326842 & 18885 & - & 4.93 & 2 & - & 1.91 $\pm$ 0.1 & - & - & - & H \\
204350437 & 218594 &  88 Aqr  & 2.62 & 2 & - & - & - & - & 2.08 $\pm$ 0.85 & H \\
374828367 & 22796 &  12 Tau  & 4.74 & 2 & - & - & - & - & 2.29 $\pm$ 0.64 & H \\
374860864 & 40409 &  36 Dor  & 3.75 & 22 & - & - & 1.12 $\pm$ 0.25 & - & - & L \\
382393070 & 18322 &  3 Eri  & 2.91 & 1 & - & 1.43 $\pm$ 0.22 & - & - & 1.35 $\pm$ 0.48 & H \\
422432907 & 115202 &  57 Vir  & 4.31 & 1 & - & 1.41 $\pm$ 0.08 & 1.26 $\pm$ 0.25 & - & - & L \\
436243682 & 37160 &  40 Ori  & 3.21 & 1 & - & 1.07 $\pm$ 0.04 & - & - & - & L \\
    	\bottomrule
    \end{tabularx}
    \begin{tablenotes}
	   \item {Notes: A machine-readable version of the table for all the 451 stars is available as online downloadable material and from CDS.}
	   \item[a]{Source: \citet{mortier_new_2013}}
	   \item[b]{Source: \citet{jofre_stellar_2015}}
	   \item[c]{Source: \citet{ppps}}
	   \item[d]{Source: \citet{ghezzi_retired_2018}}
	   \item[e]{Source: \citet{stock_evoltime_2018}}
	   \item[f]{Stars are categorised as low- (L) and high- (H) luminosity stars based on the criteria outlined in Sec.~\ref{secn:tarsel}}
	\end{tablenotes}
	\end{threeparttable}
\end{table*}

To distinguish between what can be viewed as subgiants/early red giant branch stars (low-luminosity sample), akin to the retired A-star sample, and the more evolved red giant branch and red clump stars (high-luminosity sample), we define the stars with absolute magnitude $M_{V} <$ 1.25 and colour $B-V >$ 0.8 to be high-luminosity stars (represented by the dashed black line in Fig.~\ref{fig:cmd}). The remaining stars are considered low-luminosity stars. Of the 451 stars in our sample, we find 283 high-luminosity and 168 low-luminosity stars based on the criteria above. A spectroscopic-source-wise breakdown of the low- and high-luminosity stars is provided in Table~\ref{table1:osci_stats}. We also mark individual stars as a low- or high-luminosity stars in Table~\ref{table2:spec_masses} (Column 11).

The selected targets had at least $\sim$ 27 days (one sector) of TESS data available\footnote{The stars close to the ecliptic plane ($|\beta| \lesssim 6\deg$) are not in our sample as TESS did not observe this section during the prime mission.}. The targets were observed in a 2-minute cadence, and the data was downloaded from the TESS Asteroseismic Science Operations Center (TASOC) website\footnote{\url{https://tasoc.dk/}}. For our analysis, we used the corrected light curves (PDCMAP) provided by the TESS Science Processing Operations Center (SPOC) \citep{jenkins_SPOC_2016}, which uses a pipeline similar to the Kepler Science Operations Center (KSOC) pipeline. Table~\ref{table2:spec_masses} (Columns 4 and 5) lists the \tess\ magnitudes and the observing coverage of our targets.

We first excluded any known anomalies (e.g., due to cosmic rays or stray light) by only using data with a zero pixel-quality flag as marked by the SPOC pipeline. The resulting light curves were subjected to a highpass filter of $\sim$3 \muhz\ ($\sim$4 days) to remove any slow trends. For stars 24 Sex and HD 185351, the quality of the time series was poor and showed trends with periods less than $\sim$4 days, which interfered with the detection of oscillations. Hence, we used a $\sim$50 \muhz\ highpass filter for these two stars. We then performed a 4-$\sigma$ clipping to remove significant outliers from the light curves. For seven stars, we also removed any strong features (e.g., dips, flares) to prevent interference with the oscillation detection before applying a highpass filter. To do so, we removed any sudden changes in the mean with amplitudes larger than 4-$\sigma$ of the lightcurve and of duration $\sim$0.1--2.5 days. Lastly, we combined the light curves from different sectors by normalizing the flux by dividing out the median on a per-sector basis with the timestamps preserved.

\begin{figure}
	\includegraphics[height=\textheight,width=\columnwidth, keepaspectratio, trim={0 4cm 0 7cm},clip]{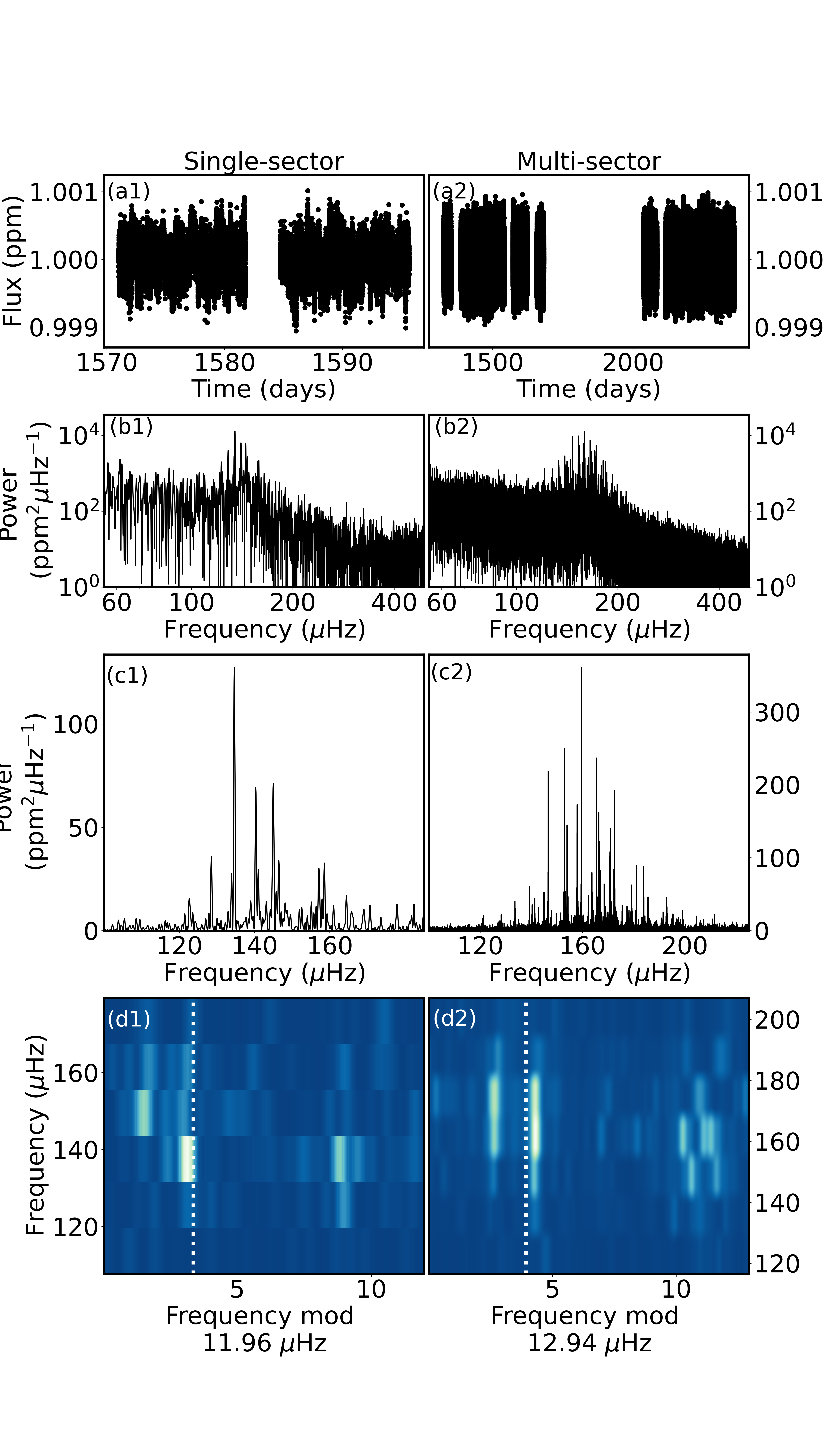}
    \caption{Comparison of the SYD results for single-sector (Column 1) and multi-sector (Column 2) data. In Column 1, we present the SYD results for the single-sector 57 Vir data, while Column 2 represents the SYD results for multi-sector 36 Dor data (22 sectors). Row a (Panels a1 and a2) depicts the light curve after combining the multi-sector data (where applicable), removing any known anomalies, and applying a highpass filter with a cut-off frequency of $\sim$3 \muhz\ ($\sim$4 days)  and a 4-$\sigma$ clipping. Row b (Panels b1 and b2) shows the power density spectrum of the corresponding star, and Row c (Panels c1 and c2) depicts the background-corrected power spectrum. Row d (Panels d1 and d2) shows the \echelle\ diagram computed from the smoothed power density spectrum, plotted using the \textit{echelle} module \citep{danhey_echelle}.
    }
    \label{fig:syd_result}
\end{figure}

Figs.~\ref{fig:syd_result}a1 and a2 show the final light curves for 57 Vir and 36 Dor, representative of short and long time series, which were observed by \tess\ for 1 and 22 sectors, respectively. We use these stars as examples in the following sections.

\section{Extraction of global seismic parameters}\label{secn:seismo}

Similar to \citet{malla_retdA_2020}, we computed the power density spectra of each light curve using a discrete Fourier transform. Figs.~\ref{fig:syd_result}b1 and b2 show the power density spectra for 57 Vir and 36 Dor. To measure the frequency of maximum acoustic power \numax\ and the large frequency of separation \dnu, we used the SYD pipeline\footnote{
Although the TESS data (sector 8) of 24 Sex has been previously used by \citet{malla_retdA_2020}, we include this star in our ensemble to include all the available data during the first three years of the prime mission and perform a homogeneous analysis using the SYD pipeline.} \citep{huber_automated_2009}. The SYD pipeline corrects for the granulation background, locates \numax\ and measures \dnu\ from the autocorrelation of the power spectrum. Figs.~\ref{fig:syd_result}c1 and c2 show the residual power spectra after the granulation background is divided out. The \numax\ and \dnu\ values are tabulated in Table~\ref{table4:obs_params} (Columns 5 and 6). 
\noindent\setlength\tabcolsep{4pt}
\noindent\setlength\tabcolsep{4pt}
\begin{table*}
	\centering
	\caption{Observed parameters and TIC-based seismic mass results for our targets }
 	\label{table4:obs_params}
	\begin{threeparttable}
	\begin{tabularx}{\linewidth}{lllc*{4}{>{\centering\arraybackslash}X}}
		\toprule
        TIC ID  & HD & Name&\teff\ & $\mathrm{\nu}_{\mathrm{max,obs,SYD}}$ & $\mathrm{\Delta\nu}_{\mathrm{obs,SYD}}$ & Source of Parallax \\
         & & & [K] &[\muhz] & [\muhz] & \\
         (1) & (2) & (3)& (4)\tnote{a} & (5) & (6) & (7)\\
         \midrule
612908 & 30856 & - & 4895 & 167.59 $\pm$ 1.96 & 13.64 $\pm$ 0.05 & Gaia eDR3 \\
1713457 & 90043 &  24 Sex  & 5027 & 191.34 $\pm$ 4.94 & 13.88 $\pm$ 0.11 & Gaia eDR3 \\
5630694 & 103616 & - & 5009 & 416.04 $\pm$ 6.63 & 30.54 $\pm$ 0.16 & Gaia eDR3 \\
6013637 & 115659 &  46 Hya  & 5136 & 58.42 $\pm$ 1.71 & nan $\pm$ nan\tnote{c} & \textit{Hipparcos} \\
11557059 & 7931 & - & 4796 & 160.49 $\pm$ 3.46 & 13.17 $\pm$ 0.09 & Gaia eDR3 \\
29921672 & 10011 & - & 5025 & 212.34 $\pm$ 5.21 & 15.19 $\pm$ 0.20 & Gaia eDR3 \\
32550970 & 8407 & - & 4922 & 101.15 $\pm$ 1.93 & 5.25 $\pm$ 0.19 & Gaia eDR3 \\
47336943 & 215049 & - & 5108 & 470.78 $\pm$ 12.48 & 29.30 $\pm$ 0.09 & Gaia eDR3 \\
48401372 & 47562 & - & 4976 & 133.48 $\pm$ 4.41 & 11.13 $\pm$ 0.11 & Gaia eDR3 \\
53873088 & 5608 & - & 4877 & 157.49 $\pm$ 15.56 & 12.77 $\pm$ 0.48 & Gaia DR2 \\
141326842 & 18885 & - & 4725\tnote{g} & 46.62 $\pm$ 0.87 & 4.40 $\pm$ 0.20\tnote{b} & Gaia DR2 \\
204350437 & 218594 &  88 Aqr  & 4439 & 10.31 $\pm$ 0.69 & 1.11 $\pm$ 0.41 & \textit{Hipparcos} \\
374828367 & 22796 &  12 Tau  & 5014 & 58.36 $\pm$ 2.34 & 5.43 $\pm$ 0.12 & Gaia DR2 \\
374860864 & 40409 &  36 Dor  & 4806 & 161.84 $\pm$ 1.08 & 12.94 $\pm$ -0.05\tnote{b} & \textit{Hipparcos} \\
382393070 & 18322 &  3 Eri  & 4706\tnote{g} & 40.14 $\pm$ 1.15\tnote{d} & nan $\pm$ nan & \textit{Hipparcos} \\
422432907 & 115202 &  57 Vir  & 4889 & 139.50 $\pm$ 1.81 & 11.96 $\pm$ -0.11\tnote{b} & \textit{Hipparcos} \\
436243682 & 37160 &  40 Ori  & 4826\tnote{g} & 41.73 $\pm$ 1.01 & 4.20 $\pm$ 0.55\tnote{c} & \textit{Hipparcos} \\
    \bottomrule
	\end{tabularx}
	\begin{tablenotes}
        \item Note: A machine-readable version of the table for all the 249 oscillating stars is available as online downloadable material and from CDS. 
	    \item[a]{Source: TESS Input Catalogue \citep{stassun_tic_2018}. We adopted an uncertainty of 100 K in the effective temperature \citep{thygesen_dteff_2012}.}
	    \item[b]{\dnu\ found using the \textit{echelle} module.}
	    \item[c]{Tentative detection.}
	    \item[d]{No \dnu\ detection.}
	\end{tablenotes}
	
	\end{threeparttable}
\end{table*}

\noindent\setlength\tabcolsep{4pt}
\vspace{50pt}
\noindent\setlength\tabcolsep{12pt}


\begin{table*}
    
	\caption{Seismic masses for our targets derived using spectroscopic parameters}
 	\label{table5:seis_from_spec}
	\begin{threeparttable}
	\begin{tabularx}{\textwidth}{lllc*{6}{>{\centering\arraybackslash}X}}
		\toprule

        TIC ID & HD & Name &L & M$_\mathrm{1}$ & M$_\mathrm{2}$ & M$_\mathrm{3}$ &  Spectroscopic source used\\
         & & &[\lsun]& [\msun] & [\msun] & [\msun] & \\
         (1) & (2) & (3) & (4) & (5) & (6) &(7)&(8) \\
         \midrule
612908 & 30856 & - & 11.38 $\pm$ 0.32 & 1.05 $\pm$ 0.08 & 0.99 $\pm$ 0.04 & 1.18 $\pm$ 0.04 & G18 \\
 &  &  & 11.49 $\pm$ 0.32 & 1.08 $\pm$ 0.08 & 1.03 $\pm$ 0.04 & 1.17 $\pm$ 0.04 & J15 \\
 &  &  & 11.35 $\pm$ 0.32 & 1.04 $\pm$ 0.08 & 0.98 $\pm$ 0.04 & 1.18 $\pm$ 0.04 & M13 \\
1713457 & 90043 &  24 Sex  & 13.33 $\pm$ 0.36 & 1.36 $\pm$ 0.11 & 1.23 $\pm$ 0.05 & 1.67 $\pm$ 0.14 & S18 \\
 &  &  & 13.18 $\pm$ 0.36 & 1.32 $\pm$ 0.1 & 1.16 $\pm$ 0.05 & 1.7 $\pm$ 0.14 & S18 \\
 &  &  & 13.05 $\pm$ 0.35 & 1.28 $\pm$ 0.1 & 1.1 $\pm$ 0.05 & 1.73 $\pm$ 0.14 & G18 \\
5630694 & 103616 & - & 5.43 $\pm$ 0.25 & 1.08 $\pm$ 0.09 & 1.25 $\pm$ 0.09 & 0.8 $\pm$ 0.04 & G18 \\
6013637 & 115659 &  46 Hya  & 97.54 $\pm$ 30.94 & 2.78 $\pm$ 0.91 & - & - & J15 \\
 &  &  & 93.14 $\pm$ 29.54 & 2.8 $\pm$ 0.91 & - & - & M13 \\
11557059 & 7931 & - & 10.18 $\pm$ 0.29 & 1.0 $\pm$ 0.08 & 0.94 $\pm$ 0.04 & 1.12 $\pm$ 0.08 & J15 \\
 &  &  & 10.19 $\pm$ 0.29 & 1.0 $\pm$ 0.08 & 0.95 $\pm$ 0.04 & 1.11 $\pm$ 0.08 & G18 \\
29921672 & 10011 & - & 10.11 $\pm$ 0.45 & 1.06 $\pm$ 0.09 & 0.84 $\pm$ 0.06 & 1.69 $\pm$ 0.15 & J15 \\
32550970 & 8407 & - & 19.64 $\pm$ 0.58 & 1.08 $\pm$ 0.09 & 0.33 $\pm$ 0.03 & 11.85 $\pm$ 1.84 & S18 \\
47336943 & 215049 & - & 5.13 $\pm$ 0.15 & 1.15 $\pm$ 0.09 & 1.04 $\pm$ 0.05 & 1.4 $\pm$ 0.11 & S18 \\
48401372 & 47562 & - & 15.7 $\pm$ 0.44 & 1.09 $\pm$ 0.09 & 0.98 $\pm$ 0.05 & 1.36 $\pm$ 0.15 & S18 \\
53873088 & 5608 & - & 13.1 $\pm$ 0.36 & 1.16 $\pm$ 0.15 & 1.12 $\pm$ 0.1 & 1.26 $\pm$ 0.42 & W16 \\
 &  &  & 13.21 $\pm$ 0.37 & 1.19 $\pm$ 0.15 & 1.16 $\pm$ 0.1 & 1.25 $\pm$ 0.42 & G18 \\
141326842 & 18885 & - & 61.17 $\pm$ 12.99 & 1.75 $\pm$ 0.39 & 1.57 $\pm$ 0.52 & 2.17 $\pm$ 0.41 & W16 \\
204350437 & 218594 &  88 Aqr  & 284.2 $\pm$ 73.95 & 2.33 $\pm$ 0.65 & 1.57 $\pm$ 1.31 & 5.13 $\pm$ 7.65 & J15 \\
374828367 & 22796 &  12 Tau  & 63.64 $\pm$ 15.86 & 1.97 $\pm$ 0.52 & 1.93 $\pm$ 0.73 & 2.05 $\pm$ 0.31 & G18 \\
374860864 & 40409 &  36 Dor  & 9.68 $\pm$ 0.4 & 0.93 $\pm$ 0.08 & 0.79 $\pm$ 0.05 & 1.28 $\pm$ 0.03 & J15 \\
382393070 & 18322 &  3 Eri  & 54.11 $\pm$ 16.65 & 1.4 $\pm$ 0.45 & - & - & M13 \\
 &  &  & 54.45 $\pm$ 16.75 & 1.41 $\pm$ 0.45 & - & - & S18 \\
422432907 & 115202 &  57 Vir  & 12.01 $\pm$ 0.41 & 0.97 $\pm$ 0.08 & 0.91 $\pm$ 0.05 & 1.1 $\pm$ 0.06 & J15 \\
 &  &  & 12.69 $\pm$ 0.43 & 1.12 $\pm$ 0.09 & 1.18 $\pm$ 0.06 & 1.03 $\pm$ 0.05 & S18 \\
436243682 & 37160 &  40 Ori  & 30.34 $\pm$ 1.0 & 0.75 $\pm$ 0.06 & 0.49 $\pm$ 0.13 & 1.81 $\pm$ 0.96 & S18 \\

 		\bottomrule
	\end{tabularx}
	\begin{tablenotes}
	   \item Note: A machine-readable version of the table for all the 249 stars is available as online downloadable material and from CDS.
	    \item $M_\mathrm{1}$ : \numax-only based seismic mass derived using Eq.~\ref{eq:scaling}
	    \item $M_\mathrm{2}$ : \dnu-only based seismic mass derived using Eq.~\ref{eqn:dnu}
	    \item $M_\mathrm{3}$ : \numax\ and \dnu- based seismic mass derived using Eq.~\ref{eqn:dnumax}
	    \item M13 : \citet{mortier_new_2013}
        \item J15 : \citet{jofre_stellar_2015}
        \item W16 : \citet{ppps}
        \item G18 : \citet{ghezzi_retired_2018}
        \item S18 : \citet{stock_evoltime_2018}
	\end{tablenotes}
	\end{threeparttable}
\end{table*}

Of the 451 stars analyzed with SYD, we found 249 stars that exhibit solar-like oscillations -- 92 low- and 157 high-luminosity stars. Table~\ref{table1:osci_stats} (Columns 3, 5 and 7, i.e., N$_\mathrm{osc}$, N$_\mathrm{low, osc}$, and N$_\mathrm{high, osc}$) provides a spectroscopic-source-wise breakdown of the solar-like oscillators in our sample. Here we note that we only have a tentative detection of \numax for four stars, as marked in Table~\ref{table4:obs_params} (Column 5). 

To verify, and possibly correct, the \dnu\ measurements by SYD, we used \echelle\ diagrams. The \echelle\ diagrams are obtained by dividing the power density spectrum into segments of length equal to a trial \dnu\ and then stacking these segments vertically. The acoustic modes of the same degree form vertical ridges when the trial \dnu\ corresponds to the correct large frequency separation. We use the SYD \dnu\ as an initial guess. Figs.~\ref{fig:syd_result}d1 and d2 depict the \echelle\ diagrams for 57 Vir and 36 Dor respectively. However, the SYD \dnu\ did not provide good vertical alignment for nine stars (marked in Table~\ref{table4:obs_params}, Column 6). For these stars, we use the \textit{echelle}\footnote{\url{https://pypi.org/project/echelle/}} module \citep{danhey_echelle} to adjust the trial \dnu\ to the final adopted value guided by the \dnu--$\epsilon$ relation by \citet{corsaro_solar-like_2012}. In such a case, we take as our uncertainty the minimum shift in \dnu\ needed to move the vertical ridges clearly out of alignment (an illustration of this can be seen in Fig. 5 of \citealt{stello_membership_2011}). We find an average uncertainty of 0.165 \muhz\ for these nine stars. Table~\ref{table4:obs_params} (Column 6) lists the final \dnu\ observed for our targets. Despite our attempts to achieve a good vertical mode alignment using SYD pipeline and/or \textit{echelle} module, we could only make a tentative detection of \dnu\ for 28 stars. Further, we could not measure \dnu\ for seven stars. Such stars are also marked in Table~\ref{table4:obs_params} (Column 6).

\section{Determining stellar mass}\label{secn:seis_mass}

We use two asteroseismic scaling relations to determine three seismic masses: one based on \numax, one based on \dnu, and one based on \numax\ and \dnu. The \numax-based mass uses the relation between \numax\ and the acoustic cutoff frequency \citep{brown_detection_1991, kjeldsen_amplitudes_1995}. Following \citet{stello_oscillating_2008}, this leads to the following relation for mass: 
\begin{equation}
    \frac{M}{\mathrm{M}_\mathrm{\sun}} \simeq 
    \left(\frac{\nu_\mathrm{max}}{\mathrm{\nu}_\mathrm{max,\sun}}\right) \left(\frac{L}{\mathrm{L}_\mathrm{\sun}}\right) \left(\frac{T_\mathrm{eff}}{\mathrm{T}_\mathrm{eff,\sun}}\right)^{-3.5}.
	\label{eq:scaling}
\end{equation}
To be consistent with \citet{stello_asteroseismic_2017} and \citet{malla_retdA_2020}, we use $\mathrm{\nu}_\mathrm{max,\sun}$ = 3090 \muhz\ and T$_\mathrm{eff,\sun}$ = 5777K \citep{huber_automated_2009}.  

We also know that \dnu\ scales with the square root of stellar mean density \citep{ulrich1986}. Hence,  we compute a \dnu-based seismic mass as follows: 
\begin{equation}
    \frac{M}{\mathrm{\mathrm{M}_\mathrm{\odot}}} \simeq 
    \left(\frac{\Delta\nu}{\mathrm{f_\mathrm{\Delta\nu}\mathrm{\Delta\nu}_\mathrm{\odot}}}\right)^{2} \left(\frac{L}{\mathrm{L_\mathrm{\odot}}}\right)^\mathrm{1.5} \left(\frac{T_\mathrm{eff}}{\mathrm{T}_\mathrm{eff, \odot}}\right)^\mathrm{-6}.
    \label{eqn:dnu}
\end{equation}

Here $f_\mathrm{\Delta\nu}$ is a necessary correction factor applied to the scaling relation linking \dnu\ and density. We use the software \textit{asfgrid}\footnote{https://ascl.net/1603.009} by \citet{sharma_2016_dnucorr} to obtain the correction factor. 

By combining Eq.~\ref{eqn:dnu} with Eq.~\ref{eq:scaling}, we measure a seismic mass using both \numax\ and \dnu\ with very little dependence on \teff\ and eliminating the need for a luminosity determination, which for data with good \dnu\ measurements, like those from \textit{Kepler}, has been the most common way to determine seismic masses on large ensembles (see \citealt{kallinger_oscillating_2010}, \citealt{apokasc_2014}, \citealt{apokasc1_2017}, \citealt{apokasc2_2018} and \citealt{yu2018asteroseismology}): 
\begin{equation}
    \frac{M}{\mathrm{\mathrm{M}_\mathrm{\odot}}} \simeq \left(\frac{\nu_\mathrm{max}}{\mathrm{\nu_\mathrm{max,\odot}}}\right)^3\left(\mathrm{\frac{\Delta\nu}{f_\mathrm{\Delta\nu}\Delta\nu_\mathrm{\odot}}}\right)^{-4}\left(\frac{T_\mathrm{eff}}{\mathrm{T}_\mathrm{eff, \odot}}\right)^\mathrm{1.5}.
    \label{eqn:dnumax}
\end{equation}

To perform a like-for-like comparison between the spectroscopic-based isochrone masses from the literature and our seismic masses, we adopt the spectroscopic parameters (\teff\ and [Fe/H]) determined from each spectroscopic survey to derive the seismic masses using the seismic scaling relations (Eqs. \ref{eq:scaling}, \ref{eqn:dnu} and \ref{eqn:dnumax}).

We used \textit{isoclassify}\footnote{\url{https://github.com/danxhuber/isoclassify}} \citep{huber_asteroseismology_2017, berger_isoclassify_2020} to compute luminosities for the solar-like oscillators in our sample. We used the spectroscopic parameters (\teff\ and [Fe/H]) from the corresponding spectroscopic source (M13, J15, W16, G18 or S18). We also adopted Gaia eDR3 parallaxes \citep{gaia_mission_2016, vallenari_gaia_edr3_2022} for stars in the magnitude range 6 < G < 21 (57 stars), Gaia DR2 parallaxes \citep{gaia_dr2} for stars with 5 $\leq$ G $\leq$ 6 (97 stars) and Hipparcos parallaxes for the rest \citep[95 stars]{leeuwen_hipparcos_2007}. The source of parallax used for each star is marked in Table~\ref{table4:obs_params} (Column 7). We discuss our choice of parallaxes in more detail in Appendix~\ref{section:parallaxes}. We also used the Tycho V$_\mathrm{T}$ photometry as an input. For three stars (NGC 4349 127, BD+20 274 and BD-13 2130), the Tycho V$_\mathrm{T}$ magnitudes were unavailable and hence, we used the 2MASS K$_s$ magnitudes. We set the \textit{dustmap} parameter to `allsky' to use a combination of reddening maps (\citealt{drimmel_three-dimensional_2003}, \citealt{marshall_modelling_2006}, \citealt{green_three-dimensional_2015} and \citealt{ bovy_galactic_2016}) to provide full sky coverage as implemented in the \textit{mwdust} package by \citet{bovy_galactic_2016}. Table~\ref{table5:seis_from_spec} (Column 4) shows the resulting luminosities using the spectroscopic parameters from each spectroscopic source. 

Now, with all the input for the scaling relations at hand, we derived the three seismic masses $M_\mathrm{1}$, $M_\mathrm{2}$ and $M_\mathrm{3}$ from Eqs.~\ref{eq:scaling}, \ref{eqn:dnu} and \ref{eqn:dnumax} using the spectroscopic parameters from M13, J15, W16, G18 and S18. We list the seismic masses in Table~\ref{table5:seis_from_spec} (Columns 5--7) for each spectroscopic source. We will compare these seismic masses to their corresponding spectroscopic masses in Section~\ref{secn:retdA}. 

\section{Mass comparison}\label{secn:retdA}

\subsection{Mass-dependent mass offset in low-luminosity stars}

There is clear evidence in the literature that the different seismic masses based on the scaling relations in Eqs.~\ref{eq:scaling}, \ref{eqn:dnu}, and \ref{eqn:dnumax} do not perform equally well. Based on analyses of nearly equal-mass cluster red giants, both \citet{miglio2016} and \citet{maddy_deltanu_2022} found that the \numax-based relation (Eq~\ref{eq:scaling}) provides results that are more in line with astrophysical expectations and that show the least scatter (among equal-mass stars). We therefore expect Eq.~\ref{eq:scaling} to be the best seismic mass to use for our comparison with the spectroscopic results. However, we first wished to check for our sample that Eq.~\ref{eq:scaling} provides the most robust results. Because our sample is comprised of field stars, we do not have an astrophysical 'true' mass for each star to serve as a reference. However, we can still compare our three different seismic masses to the spectroscopic mass, which we can use as our reference mass even if it has a suspected offset from the true mass. This is because the potential offset is expected to change smoothly with mass in an ensemble sense. Hence, the scatter in the differences between each seismic mass scale and the spectroscopic one, for a given reference mass, still gives an indication of the overall precisions of the different seismic mass scales.

\begin{figure*}
	\includegraphics[width=\textwidth,height=0.9\textheight,keepaspectratio, trim={0 2cm 0 0}, clip]{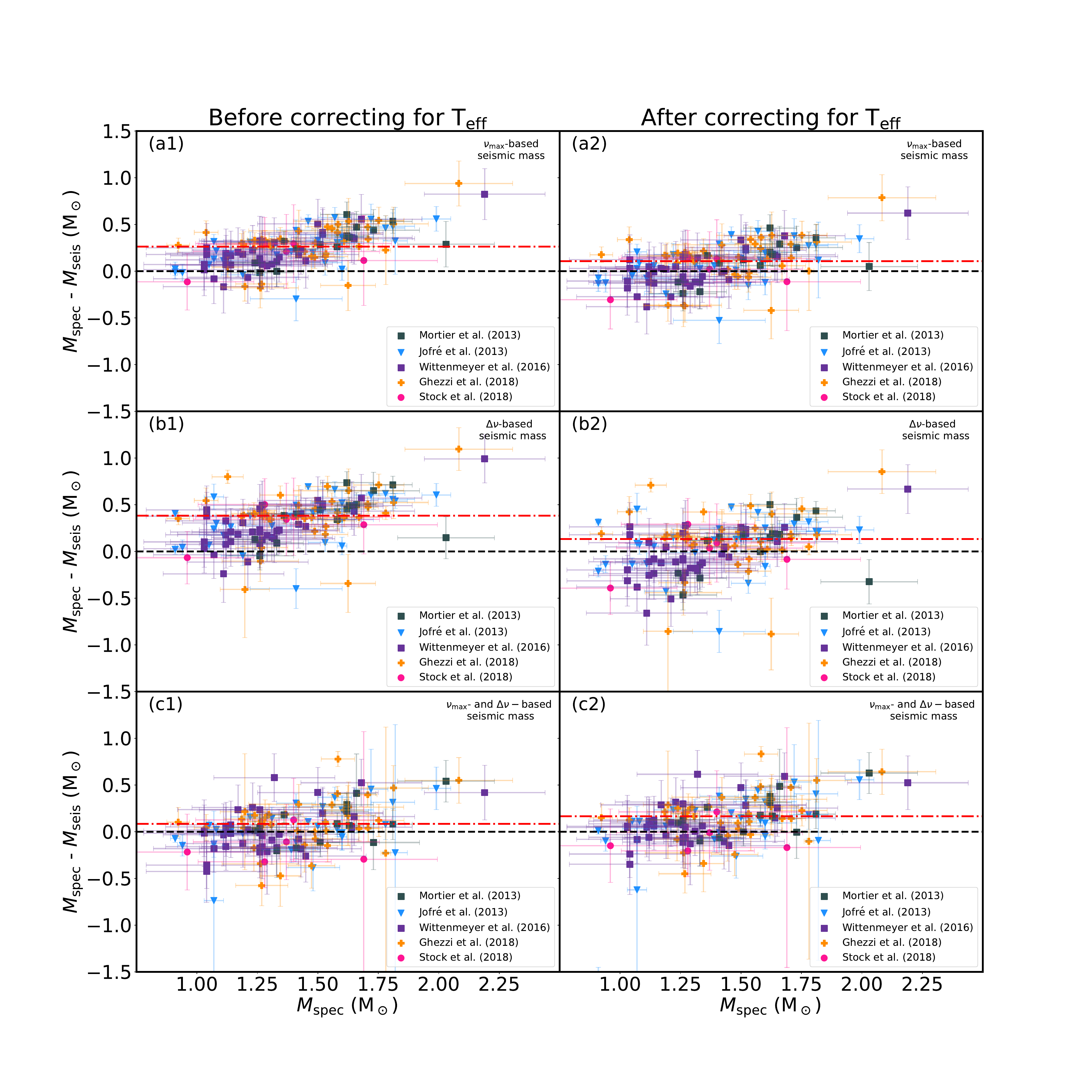}
    \caption{Difference between the spectroscopic and seismic mass plotted as a function of spectroscopic mass for 92 low-luminosity stars with solar-like oscillations. Column 1 depicts the mass offset before any corrections were made, and Column 2 depicts the mass offset after correcting the offset in the weighted average by decreasing the \teff\ by 100K. The seismic masses used in panels a1 and a2 are the \numax-based seismic masses, in b1 and b2 are the \dnu-based seismic masses, and in c1 and c2 are the ‘\numax\ + \dnu’- based seismic masses. The green triangles use the spectroscopic masses and parameter by M13 (14 stars), blue squares use those by J15 (31 stars), purple diamonds use those by W16 (36 stars), orange pluses use those by G18 (40 stars) and pink circles those by S18 (5 stars). The dashed black line represents no difference between the spectroscopic and seismic masses. The dash-dotted red line represents the weighted average mass offset.}
    \label{fig:mass_diff}
\end{figure*}

Of the 168 low-luminosity stars, 92 show solar-like oscillations (Table~\ref{table1:osci_stats}, columns 4 and 5). We plot the difference between the seismic and spectroscopic mass scales as a function of the spectroscopic mass in Fig.~\ref{fig:mass_diff} (Panels a1, b1 and c1, with a panel for each seismic scale). Here, we use all five different spectroscopic sources.  The 22 low-luminosity stars listed in Table~\ref{table2:spec_masses} appear in more than one of our spectroscopic sources and are plotted multiple times. Stars with tentative detections are also included in this analysis. In agreement with \citet{miglio2016} and \citet{maddy_deltanu_2022}, we indeed see the smallest scatter (at fixed reference mass) for the \numax-based seismic mass (Eq.~\ref{eq:scaling}), supporting its use. In addition, we note that the three seismic mass scales have different zero-points relative to the spectroscopic reference, but interestingly, that difference is largely proportional to the exponent of \teff\ in the three mass scaling relations. Hence, it seems this is not dominated by an intrinsic zero-point difference between the seismic scales, but more likely pointing to a systematic offset in the adopted \teff\ scale. If we adjust the \teff\ scale by -120 K, we can bring all three seismic mass scales to a similar zero-point\footnote{The spectroscopic masses remain unchanged and were not adjusted based on the adjusted \teff\ scale.} (Fig.~\ref{fig:mass_diff}a2, b2 and c2). We note that this \teff\ shift is comparable to our adopted \teff\ uncertainty (100 K, \citealt{thygesen_dteff_2012})\footnote{Because the three mass scaling relations also depend differently on luminosities, we did try to reconcile the average mass offsets by changing luminosity under the assumption that there is no parallax offset or \teff-offset. However, this required increasing luminosity by 0.5 \lsun, which is beyond what we expect from the systematic parallax errors \citep{lindegren_gaia_2018, zinn_radius_2019, gaia_dr2}.}.  

\begin{figure*}
	\includegraphics[width=\textwidth, keepaspectratio, trim={0cm 0cm 0cm 0cm}, clip]{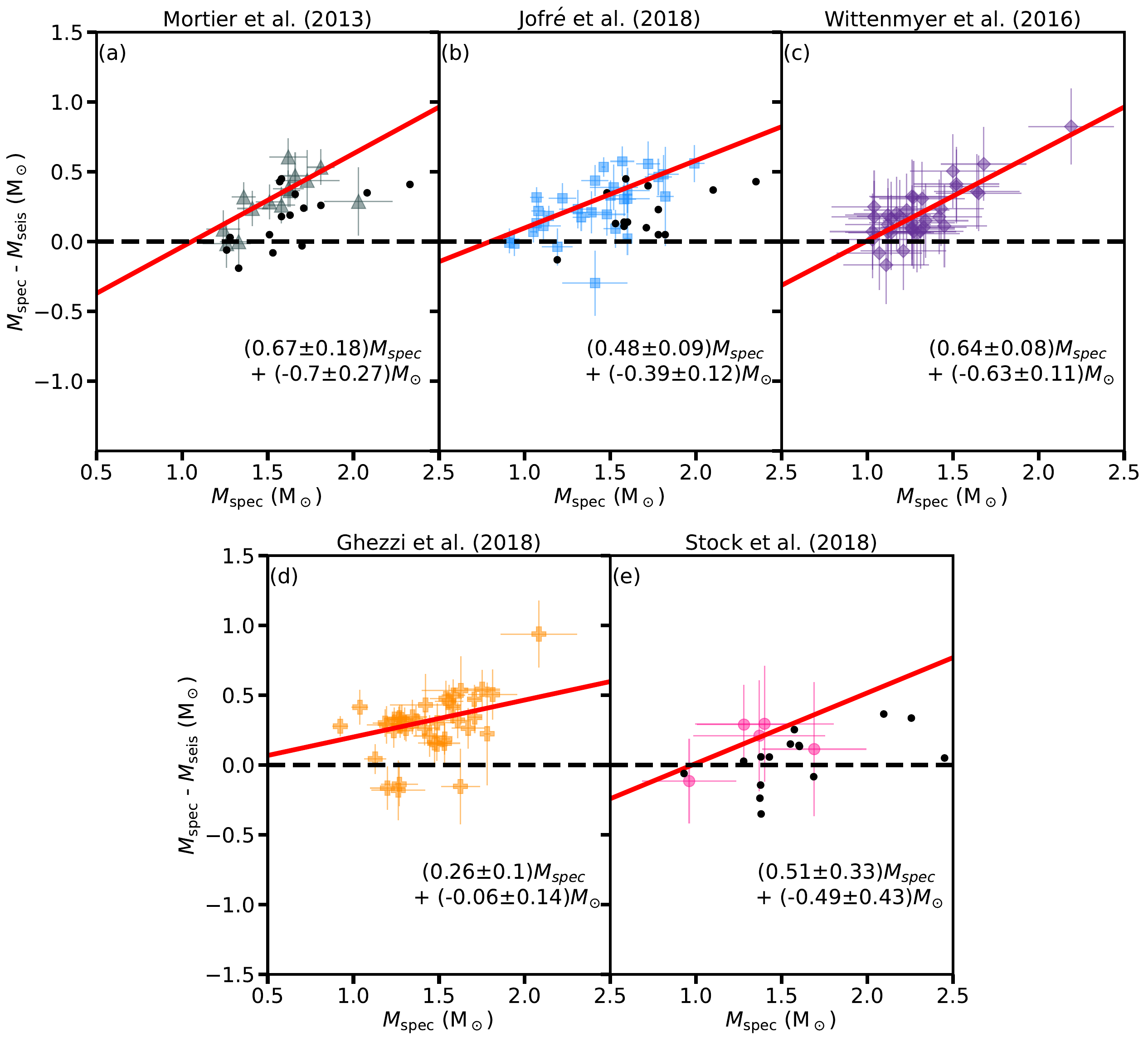}
    \caption{First order polynomial fit to the difference between the spectroscopic and seismic mass plotted as a function of spectroscopic mass for 92 low-luminosity stars with solar-like oscillations. Panel (a) uses the spectroscopic masses and parameters by M13 (14 stars), (b) uses those by J15 (31 stars), (c) uses those by W16 (36 stars), (d) uses those by G18 (40 stars) and (e) uses those by S18 (5 stars) We used the \numax-based seismic masses (\mnumax) without any \teff-correction applied. The black circles in Panels a, b and e refer to the results by \citet{malla_retdA_2020}. The solid red line represents an uncertainty-weighted least-squares fit to the data. The dashed black line represents no difference between the spectroscopic and seismic masses. Here, we used all five different spectroscopic sources.}     
    \label{fig:mass_diff_indivi}
\end{figure*}

For our main comparison between seismic and spectroscopic masses, we adopted the \numax-based seismic scale. However, we did not apply the 120K \teff\ shift in the following as we do not claim to know what the true mass scale of our sample is. In Fig.~\ref{fig:mass_diff_indivi}, we show this comparison for each spectroscopic source. By plotting our results together with the ensemble study results by \citet{malla_retdA_2020} for M13, J15 and S18 (black dots), we find our results are consistent with their findings. Further, from Fig.~\ref{fig:mass_diff_indivi}, we find that the mass offset is the smallest for the S18 subsample in which evolutionary speeds were taken into account.

\subsubsection{Fitting the mass-dependent mass offset}
\begin{figure}
	\includegraphics[width=\columnwidth, keepaspectratio, trim={0cm 0cm 0cm 0cm}, clip]{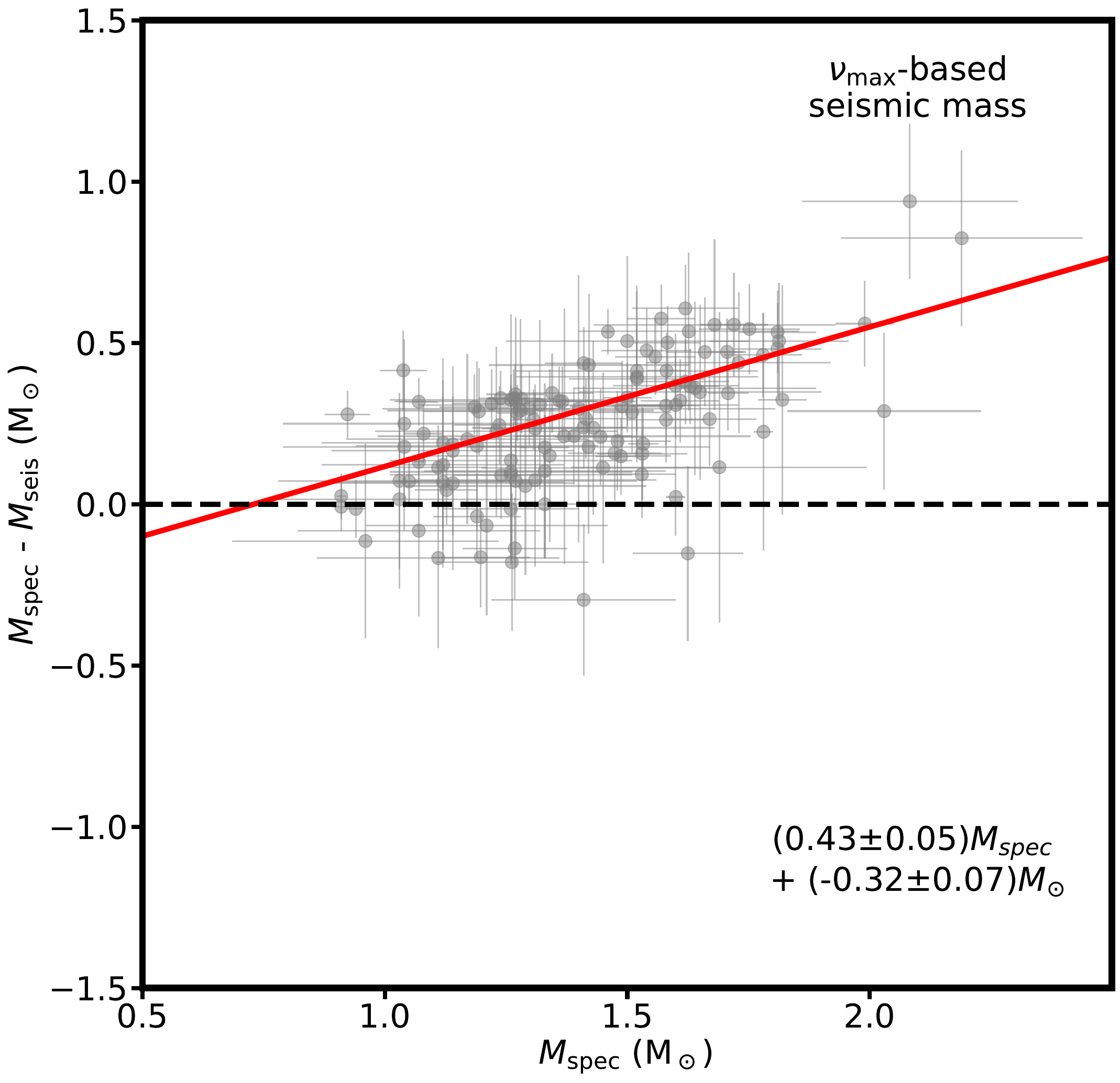}
    \caption{First order polynomial fit to the difference between the spectroscopic and seismic mass plotted as a function of spectroscopic mass for 92 low-luminosity stars with solar-like oscillations. Here, we used the \numax-based seismic masses (\mnumax) without any \teff-correction applied. The solid red line represents an uncertainty-weighted least-squares fit to the data. The dashed black line represents no difference between the spectroscopic and seismic masses. Here, we used all five different spectroscopic sources. The 22 low-luminosity stars that appear in more than one of our spectroscopic sources appear multiple times.}     
    \label{fig:bic}
\end{figure}

It is clear from Fig.~\ref{fig:mass_diff_indivi} that there is a mass-dependent mass offset. To quantify this offset, we fitted polynomials of varying orders and a step function (following a suggestion by \citealt{malla_retdA_2020}) to the mass offset without \teff-correction and computed the Bayesian Information Criterion (BIC). \citet{schwarz_bic_1978} first proposed BIC as a way to choose between alternate models in regression where the model with the lowest BIC is the preferred. We used the weighted least squares (WLS) method from the \textit{statsmodel} module in python to fit polynomials of order 0--7 and the step function. We found the BIC to be minimum for a first-order polynomial fit (BIC = -82.89) of the following form:
\begin{equation}
    M_\mathrm{spec} - M_\mathrm{seis} = (0.43 \pm 0.05)M_\mathrm{spec} + (-0.32 \pm 0.07) M_\mathrm{\odot}.
    \label{eqn:mass_offset_fit}
\end{equation}
However, a second-order polynomial fit had a BIC of -78.58, with the difference in BIC ($\Delta$BIC) being 4.31. This value of $\Delta$BIC implies that the first-order polynomial is a good fit to our mass offset\citep{fabozzi_bic_2014, raferty_1995_bic}. However, we cannot completely rule out a second-order polynomial. Future work is needed to expand beyond the limits of the current study and comprehensively model the offset to potentially rule one out. Fig.~\ref{fig:bic} shows how well the first order polynomial model fits the mass offsets. We advise caution on the use of Eq.~\ref{eqn:mass_offset_fit} because it is only applicable to stars with spectroscopic masses 0.91--2.90 \msun, as can be inferred from Table~\ref{table1.5:spec_mass_range} (Column 5). 

From Fig.~\ref{fig:mass_diff_indivi} (Panels a--e), we note that fitting the first order polynomial to individual spectroscopic sources yields intercepts between -0.7 \msun\ and -0.06 \msun\ and slopes between 0.26 and 0.67. We find that the slopes for individual fits of M13 (0.67 $\pm$ 0.18), J15 (0.48 $\pm$ 0.05) and S18 (0.51 $\pm$ 0.33) are consistent with each other. G18 has the least slope of 0.26 $\pm$ 0.1 owing to the larger scatter in its mass offsets. The W16 sample has the most precise fit with a slope of 0.64 $\pm$ 0.08 while S18 has the least precise fit because of only five data points.\footnote{It follows from Eq.~\ref{eqn:mass_offset_fit} that the planet occurrence-mass-metallicity correlation remains unaffected by the mass offset.}

Despite the larger ensemble size used in this work, our study still suffers from a deficit of low-luminosity stars with spectroscopic masses below 1 \msun\ and above 1.8 \msun\footnote{Although the \citet{malla_retdA_2020} ensemble has three stars with spectroscopic masses greater than 2 \msun, they have not been observed by TESS during its three-year primary mission.}. From Table~\ref{table1.5:spec_mass_range} (Column 5), we note that our low-luminosity solar-like oscillators covers a spectroscopic mass range of 0.91--2.19 \msun. As a result, we are unable to define how the mass offset behaves at masses well below 1 and well above 2 \msun. With \tess\ observing more evolved stars in the late subgiant/early red giant branch regime, future work might be able to explore the mass-dependent mass offset at sub-solar masses and masses above 2 \msun. 

\subsection{The high-luminosity stars}\label{secn:high_lum}
\begin{figure}
	\includegraphics[width=\columnwidth,height=0.9\textheight,trim={0 0 0 0}, clip, keepaspectratio]{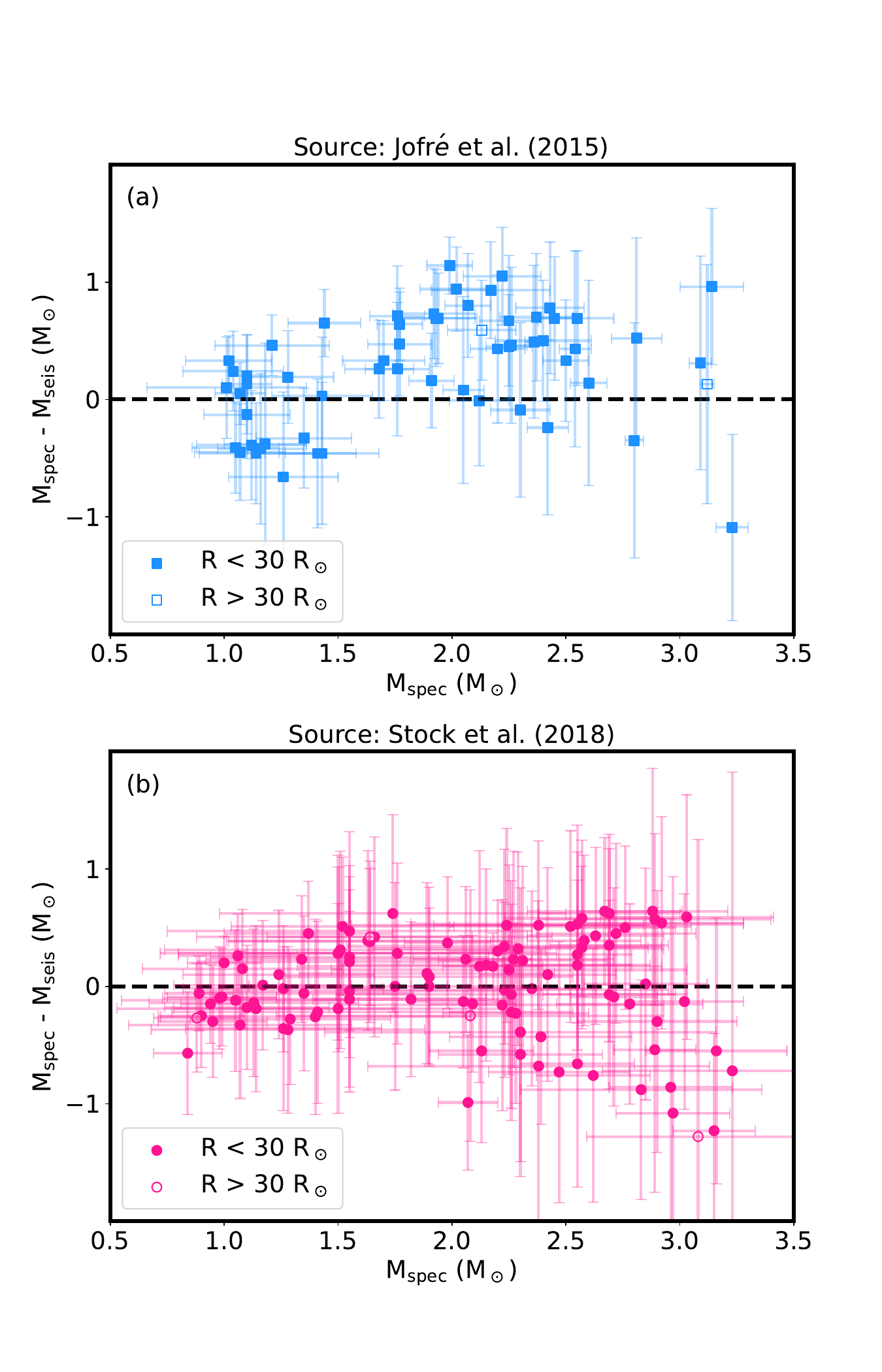}
    \caption{Difference between the spectroscopic and seismic mass plotted as a function of spectroscopic mass for 157 unique high-luminosity stars with solar-like oscillations, plotted separately according to the spectroscopic parameters used. The seismic masses used in row (1) are the \numax-based seismic masses, in row (2) are the \dnu-based seismic masses, and in row (3) are the`\numax\ + \dnu'- based seismic masses. The blue squares use the spectroscopic masses and parameters by \citet[61 stars]{jofre_stellar_2015} (Column 1), and pink circles use those by \citet[111 stars]{stock_evoltime_2018} (Column 2). Unfilled markers represent the stars with R > 30 \rsun. The black dashed line indicates zero difference between the spectroscopic and seismic masses.
    }
    \label{fig:mass_diff_indivi_high_lum}
\end{figure}

Of the 451 stars, we have 283 high-luminosity stars, according to the criteria listed in Sec.~\ref{secn:tarsel} (also marked individually in Table~\ref{table2:spec_masses}, column 11). The most luminous stars in this sample oscillate at frequencies too low to resolve with the length of our current TESS data and our highpass filter of ~3 \muhz. The most luminous star in our sample oscillates at 0.3 \muhz. We need at least 15 sectors of TESS data (i.e. more than a full year’s coverage) to detect oscillations at this \numax. However, we find no such stars with multi-sector TESS data oscillating at such small frequencies in our sample. To be conservative, we do not report any seismic detections below a \numax\ of 8 \muhz. Taking this hard detection limit into account, we have 157 solar-like oscillators, a spectroscopic source-wise breakdown of which is provided in Table~\ref{table1:osci_stats} (columns 6--7). Similar to our low-luminosity sample, we compare our seismic masses to the spectroscopic masses by plotting the difference between the two as a function of spectroscopic mass in Fig.~\ref{fig:mass_diff_indivi_high_lum}. We do not have a statistically significant high-luminosity star sample size for the spectroscopic sources M13, W16 and G18. Hence, these sources have been excluded from the figure and any discussion concerning the high luminosity stars. 

From Fig.~\ref{fig:mass_diff_indivi_high_lum}, we see a rather complicated mass offset. This observed offset is non-trivial to understand from a stellar evolution point, unlike the arguments used on the low-luminosity sample, where stellar tracks are well separated and non-overlapping. The high-luminosity stars are mostly in the region of the colour-magnitude diagram where the tracks clump together, making it harder to determine stellar mass using stellar models. Therefore, determining the stellar masses using a more model-independent method like asteroseismology might provide more robust stellar masses for these stars. However, the \numax-based seismic relation (Eq.~\ref{eq:scaling}) seems to break down for stars with radius larger than about 30-50 \rsun\ (\citealt{zinn_radius_2019} and \citealt{zinn_radius_2023}), which affects up to six stars\footnote{namely, HD 27256, 73 Tau, HD 2774, 88 Aqr, HD 11977, 19 Sex} in our sample. 

\section{Conclusions}\label{secn:conclusions}
We constructed a sample of 451 evolved stars from five different spectroscopic sources: M13, J15, W16, G18, and S18. Of the 451 stars, we found 249 solar-like oscillators from 2-min cadence TESS data and measured their \numax\ and \dnu.

To explore the mass-dependent mass offset between the spectroscopic and seismic masses for subgiants and red giants, we determined the seismic masses from three different scaling relations (Eqs.~\ref{eq:scaling}, \ref{eqn:dnu} and \ref{eqn:dnumax}) using spectroscopic parameters from M13, J15, W16, G18 and S18. We divided our initial sample into two subsets: 168 low- and 283 high-luminosity stars, such that the low-luminosity stars are synonymous with the previously studied retired A-stars (as defined by G18). Of these, we found 92 low- and 157 high-luminosity stars with solar-like oscillations. Comparing the spectroscopic and seismic masses for low-luminosity stars (mostly late subgiants and early red giants) reveals an increasing mass offset with mass between the two stellar mass scales. However, we found that the planet occurrence-mass-metallicity correlation does not significantly change when adopting the seismic masses. Despite our efforts to expand our ensemble size and to include more stars in the mass ranges not previously covered by \citet{malla_retdA_2020}, our study is deficient in low-luminosity stars with spectroscopic masses well below 1 \msun\ and above 2 \msun. A further study exploring these mass ranges may be needed to explore further the nature of the mass-dependent mass offset we observe. 

We also found a mass offset between the spectroscopic and seismic masses for high-luminosity stars (primarily consisting of the red clump and more luminous stars). A possible explanation for the observed offset is the error-prone mass determination using isochrones due to the clumping together of evolutionary tracks in this region of interest.

\section*{Acknowledgements}

D.S. is supported by the Australian Research Council (DP190100666). T.R.B. acknowledges support from the Australian Research Council through Laureate Fellowship FL220100117. This paper includes data collected by the TESS mission, which are publicly available from the Mikulski Archive for Space Telescopes (MAST). Funding for the TESS mission is provided by NASA's Science Mission directorate. SPM thanks Joel C. Zinn for helping with parallax correction.

\section*{Data Availability}
The TESS data used in this article can be downloaded from TASOC (\url{www.tasoc.dk}). All the tables in this paper are available on CDS in a machine-readable format.




\bibliographystyle{mnras}
\bibliography{ref} 



\appendix

\section{A note on use of parallaxes} \label{section:parallaxes}
For computing luminosities using \textit{isoclassify} in Sec.~\ref{secn:seis_mass}, we used parallaxes from three different sources: Gaia eDR3, DR2 and \textit{Hipparcos}.

The Gaia eDR3 catalogue\footnote{In this paper, we use Gaia eDR3 and DR3 synonymously because the Gaia DR3 builds upon the Gaia eDR3 catalogue \citep{gaia_dr3_2022}}, improves the precision by a factor of 1.5, compared to Gaia DR2 \citep{lindegren_gaia_edr3_2021}. So we prefer Gaia eDR3 parallaxes for our targets if they are available. However, despite the precision improvement, the Gaia eDR3 parallaxes suffer from systematic offsets and must be corrected for a zero-point offset. For our Gaia eDR3 parallaxes, we use the python code\footnote{\url{https://gitlab.com/icc-ub/public/gaiadr3_zeropoint}} by \citet{lindegren_gaia_edr3_2021} to compute the zero-point offset based on the position, colour and magnitude of the star. The zero-point offsets are then subtracted from the Gaia eDR3 parallaxes. However, these corrections can only be applied to stars with $6 < G < 21$. To overcome these limitations, we only use the Gaia eDR3 parallaxes for stars with $6 < G < 21$ and Gaia DR2 parallaxes for stars with $G < 6$. We corrected our Gaia DR2 parallaxes using an $\verb|astrometric_pseudocolor|$- and magnitude-based relation given by \citet{zinn2019_gaia}. 

We know that Gaia DR2 parallaxes are prone to calibration issues for stars brighter than G = 5 \citep{drimmel_gaia_errors_2019}. Hence, we used Hipparcos parallaxes for stars with G < 5. Also, we do not have any Gaia eDR3/DR2 measurement available for 66 stars. For these 66 stars, we use \textit{Hipparcos} parallaxes instead. We also use \textit{Hipparcos} parallaxes for three stars that have a larger fractional error in Gaia parallaxes than in \textit{Hipparcos}.

By comparing the corrected eDR3 parallaxes to the DR2 ones for the stars that have both measurements (107 stars, where 58 stars are solar-like oscillators and 49 stars are not), we obtain a median fractional difference of about 0.8\% between the two parallaxes measurements. For seven stars\footnote{Five of the seven stars are  solar-like oscillators, namely TIC 29921672, TIC 422279419, TIC 156291668, TIC 165193011, and TIC 5630694. Two are non-oscillating stars (TIC 439871393 and TIC 62734192).}, we find the fractional difference to be more than 5\%, but less than 32\%. A possible reason for a larger fractional difference for these stars is due to the improvement in parallaxes from Gaia DR2 to eDR3. In conclusion, we find that the parallaxes from Gaia eDR3 corrected using the code by \citet{lindegren_gaia_edr3_2021} and from Gaia DR2 corrected using the relation by \citet{zinn2019_gaia} are consistent with each other.

Table~\ref{tableB1:plx_comp} summarises our discussion on the use of parallaxes in this paper. We also provide the source of parallax for each solar-like oscillator in Table~\ref{table4:obs_params} (Column 7).  

\noindent\setlength\tabcolsep{2pt}%
\begin{table}
	\centering
	\caption{Parallax sources used for our targets}
	\label{tableB1:plx_comp}
	\begin{threeparttable}
	
	\begin{tabularx}{\columnwidth}{P{1.4cm}c*{4}{>{\centering\arraybackslash}X}} 
		\toprule
		 Source of Parallax & Criteria & Correction applied & $N_\mathrm{tot}$ & $N_\mathrm{tot, osc}$  \\
		(1) & (2) & (3) & (4) & (5)\\
		\midrule
         Gaia eDR3 & 6 < G < 21 & \citet{lindegren_gaia_edr3_2021} & 109 & 57 \\
        Gaia DR2 &  5 $\leq$ G $\leq$ 6 & \citet{zinn2019_gaia} & 157 & 97 \\
        \textit{Hipparcos} & G<5 and exceptions & N/A & 185 & 95 \\
    	\bottomrule
    \end{tabularx}
    \begin{tablenotes}
        \item {Note: Exceptions include stars who do not have Gaia parallaxes, and stars whose fractional error in Gaia parallaxes is larger than in Hipparcos.}
        \item {$N_\mathrm{tot}$: Number of stars in our entire sample for which we adopt this parallax source}
        \item {$N_\mathrm{tot, osc}$: Number of oscillating stars in our sample for which we adopt this parallax source}
        \item{G : Magnitude (\verb|phot_g_mean_mag|) from Gaia eDR3. If not available, we used the magnitudes from Gaia DR2.}
        \item {A machine-readable version of the table is available as online downloadable material and from CDS.}
    \end{tablenotes}
	\end{threeparttable}
\end{table}

\bsp	
\label{lastpage}
\end{document}